\documentstyle[a4,11pt,epsf]{article}
\newcommand{\be}{\begin{eqnarray}}
\newcommand{\ee}{\end{eqnarray}}
\newcommand{\nn}{\nonumber}
\newcommand{\ovl}{\overline}
\newcommand{\ra}{\rightarrow}

\def\0n{0\nu\beta\beta}

\begin{document}
\begin{center}
{\Large \bf Collider Signatures of Sneutrino Cold Dark Matter} \\
\vspace{0.5cm}
{\large St. Kolb, M. Hirsch$^\dagger$, H.V. Klapdor-Kleingrothaus 
and O. Panella$^{\ddagger}$} \\
\vspace{0.3cm}
{\it Max-Planck-Institut f\"ur Kernphysik,
P.O. 10 39 80, D-69029 Heidelberg, Germany} \\ \vspace{0.3cm}

{\it $^{\dagger}$ Instituto de F\'{\i}sica Corpuscular
 -- C.S.I.C., Departamento de F\'{\i}sica Te\`orica, Universitat of 
Val\`encia, 46100 Burjassot, Val\`encia, Spain} \\ \vspace{0.3cm}

{\it $^{\ddagger}$ Istituto Nazionale di Fisica Nucleare, Sezione di Perugia,
Via A. Pascoli, I-06123 Perugia, Italy}
\end{center}

\vspace{0.5cm}
\begin{abstract}
Decays of sneutrinos are considered in the case that in the
presence of lepton-number violation in the sneutrino sector 
the lighter $\tau$-sneutrino is the Lightest Supersymmetric
Particle and the Cold Dark Matter in the Universe. 
In such circumstances the signals from sparticle decays differ 
considerably from the ``standard'' case where the lightest 
neutralino is the Lightest Supersymmetric Particle and it
is found that in a wide range of parameters compatible with
the sneutrino Cold Dark Matter hypothesis signatures 
characteristic for such a scenario should be easily observable 
at for example a Next Linear Collider. 
\end{abstract}

\vspace{0.5cm} 

Recently there has been particular interest in sneutrinos,
the scalar counterparts of the neutrinos appearing in 
supersymmetric (SUSY) extensions of the Standard Model (SM) 
\cite{susy}, due to the fact that Lepton-number ($L$) violation
present in the sneutrino sector of the low-energy SUSY Lagrangian 
\cite{lviol,seesaw} has huge impacts both on sneutrino
and neutrino phenomenology. The intimate connection between sneutrino
and neutrino properties implies that if the sneutrino bears ``Majorana''
properties so does the neutrino, and vice versa. Sheding light on the
as yet unresolved questions concerning the neutrino mass \cite{massneu}
is an unexpected feature of the SUSY version of the SM
once $L$  violation by sneutrinos exists. Further
potentially observable consequences include $L$-violating
processes such as neutrinoless double beta $\0n$-decay (without the
need of $R$-parity violation) \cite{db}, $e^- e^- \ra \chi^- \chi^-$ 
and sneutrino oscillations \cite{obs,seesaw}.
 
Particularly interesting is the impact of neutrino/sneutrino
properties on Cosmology and the still unresolved questions of
nature and origin of the dark matter in the universe.
Whereas massive neutrinos are long known to
be an interesting candidate for the hot component of dark matter
in the universe, it emerged only recently that in the presence of
$L$ violation sneutrinos may well serve as cold component of dark 
matter \cite{cdm}. The big experimental efforts (for a recent 
overview see e.g. \cite{proc}) that are being made to observe e.g. 
neutrino oscillations (LSND, GNO, Super Kamiokande etc.), 
$\0n$-decay (Heidelberg-Moscow, Nemo, GENIUS among others), to 
establish SUSY (e.g. at the LHC, Next Linear Collider), to observe 
CDM (HDMS, CDMS, GENIUS among others) etc. will hopefully provide 
interesting insight into these questions soon.  

One of the benefits of SUSY extensions of the SM is that under 
certain circumstances the LSP is stable and provides a candidate
for the Cold Dark Matter in the Universe (CDM).
This possibility has been ruled out for $L$-conserving
sneutrinos: sneutrinos with masses of order $100GeV$ annihilate 
rapidly via $s$-channel $Z^0$-exchange so that no cosmologically
interesting relic abundance is obtained. To reduce the annihilation
rate it was proposed that sneutrinos should be either very light
($2GeV$ \cite{lightsneu}) or very heavy (of order $1TeV$ 
\cite{heavysneu}). The first case is excluded from $Z^0$-width
measurements, the second case is excluded from direct nuclear 
recoil experiments like the Heidelberg-Moscow setup 
(\cite{heavysneu} and ref. therein). 

The situation changes if $L$ is not conserved in the
sneutrino sector. $L$ violation for the 
$SU(2)_L$-doublet sneutrinos has been introduced in a 
model-independent way in \cite{lviol} by means of an effective
$L$-violating mass term 
$m_M^2 \tilde{\nu}_L \tilde{\nu}_L + h.c.$ which 
may be present below the electroweak symmetry breaking scale.
In \cite{seesaw} $L$ is violated by $SU(2)_L$-singlet sneutrinos 
which, below the electroweak symmetry breaking scale, mix with 
the $SU(2)_L$-doublet sneutrinos. The result of both approaches 
is that the light sneutrino states $\tilde{\nu}_i^{l,h}$ 
(``$l$'' light, ``$h$'' heavy, $i$ generation index) are no 
longer the interaction eigenstates, violate $L$ and exhibit a 
mass difference. 

$L$ violation reopens the possibility that the CDM is made
up of sneutrinos \cite{cdm}: if the mass
splitting of two sneutrino states (of the same generation) is bigger
than about $5GeV$ the sneutrino may be a viable CDM candidate due to the
fact that the coupling to the $Z^0$ is off-diagonal, {\it i.e.}
$Z^0 \tilde{\nu}_i^l \tilde{\nu}_i^h$
for any generation $i$. This is a consequence of angular 
momentum conservation and Bose symmetry, thus suppressing both the 
annihilation of sneutrinos in the early universe and the elastic
sneutrino scattering rates in counting experiments. Furthermore,
sneutrino pair annihilation into gauge bosons and neutrinos 
has to be suppressed resulting in the additional conditions
$m_{\tilde{\nu}^{LSP}} < m_W,\; M_1 > 200 GeV$ where $M_1$
is the mass of the $U(1)_Y$ gaugino.  

\begin{figure}[t]
\hspace*{2cm}
\epsfxsize95mm  
\epsfbox{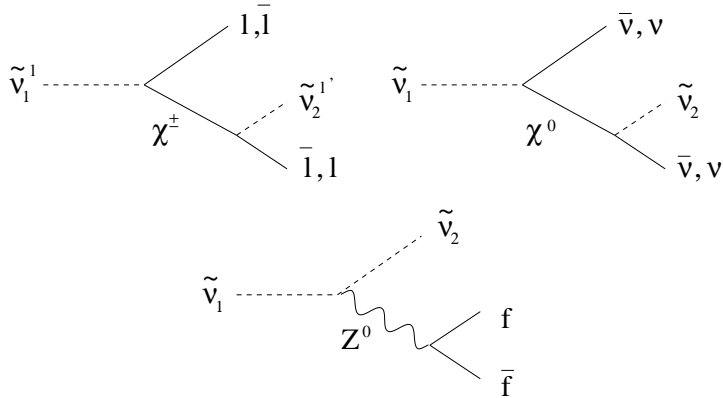}
\caption{Three-body sneutrino decays into fermions and LSP-sneutrino.}
\label{feynhl}
\end{figure}

Bounds on the sneutrino mass-splitting can be derived from
the upper limits on neutrino masses, on $\0n$-decay and from the
Baryon Asymmetry in the Universe (BAU). The constraints from
the neutrino mass \cite{lviol,seesaw} allow the sneutrino 
mass-splitting being large enough for the third generation so 
that the lighter $\tau$-sneutrino may be the CDM. On the other 
hand, the requirement that the BAU should not be erased by 
sneutrino-induced $L$-violating scatterings after the 
electroweak phase transition does not allow for sneutrino
CDM \cite{bau}.
%

However, this bound holds only in the case when 
sphaleron-mediated processes are in equilibrium immediately
after the electroweak phase transition. If {\it e.g.} the BAU is 
generated during the electroweak phase transition the lighter
$\tau$-sneutrino may be the CDM.   

If the conditions mentioned before are not fulfilled the sneutrino 
abundance alone cannot account for the CDM in the universe, but still 
the lightest sneutrino could be the LSP. This is not true in the very 
special case of the mSUGRA model. Though there are ranges in the 
$M_1,\mu,\tan\beta, m_0, A_0$ parameter space where the sneutrino
could be the LSP, the right-handed charged sleptons get too light to be 
compatible with the bounds on their masses \cite{pdg}.

The possibility of one sneutrino state being the LSP and in the
case of conserved $R$-parity being a CDM candidate may have
particularly interesting consequences for the sneutrino decay 
signatures to be 
expected at colliders. Usually the lightest neutralino is assumed to 
be the LSP and, if $R$-parity is conserved, {\it e.g.} the prevailing 
signature of a 
slepton pair produced at a Next Linear Collider (NLC) 
(see {\it e.g.} \cite{nlc} and ref. therein) is simply an acoplanar 
lepton pair and missing momentum from the neutralino LSP. The situation is
different for sneutrino LSP since the decay chains of the sparticles     
often contain heavy sneutrino states whose decays into the LSP sneutrino
produce characteristic final states, see below. Therefore in case
the lightest sneutrino being the LSP the expected SUSY signals differ
considerably from the case the neutralino being the LSP.   

In what follows we will assume that the lighter $\tau$-sneutrino is 
the LSP 
with the sneutrinos of the first two generations being almost 
degenerate in mass, and that there is a substantial mass splitting 
(of order $10GeV$) in the $\tau$-sector since this possibility is the 
cosmologically interesting one.  
The characteristic feature of such a scenario should be that
the decay chains of decaying sparticles contain copiously 
%
%
$\tau \tau$-pairs, $l \tau$-pairs ($l=e,\mu$) and jets since such 
pairs are produced by the decays of sneutrinos not being the
LSP into the LSP-sneutrino. 
%
%

On the other hand, if the lightest neutralino is the LSP 
$\tau$-pairs and jets are produced 
by transitions from non-LSP neutralinos into LSP-neutralinos.
%
%
%
%
However, the generation off-diagonal signature could be reproduced by
such decays only if generation mixing in the neutralino-slepton-lepton
vertex is sizeable. If a mixing matrix is introduced into the 
neutralino-slepton-lepton vertex the current bounds on the branching 
ratio $BR$ of transitions $\tau \ra l \gamma$ ($l$=$e,\mu$), 
$BR < {\cal O}(10^{-6})$ \cite{pdg}, set limits on its entries.
Applying mass insertion on the external $\tau$-line the relevant  
expression can be found in \cite{barbieri}. 
A rough estimate on the generation mixing may be obtained if one
assumes that there are  
no cancelations between contributions from different 
neutralino and slepton mass-states entering the amplitude.
For a common SUSY particle mass of $100 GeV$ and for
a bino-dominated neutralino an upper 
bound on the mixing matrix elements connecting the third with the 
first and second generations $|U_{l \tau}|^2 < {\cal O}(10^{-2})$ 
can be deduced.
%
%
%
%
The limit depends on the ratio $r=m_{\chi^0}^2/m_{\tilde{l}}^2$ and 
becomes slightly more restrictive for smaller values of $r$. In the limit 
where $r$ becomes very large no constraint is obtained, but since it is 
assumed that one neutralino is the LSP this case is not of interest.
%
%
Therefore it seems unlikely that generation off-diagonal lepton pairs
are produced copiously by decaying heavy neutralinos. An additional
possibility to distinguish between the two cases is provided by
the different angular distributions of off-diagonal lepton pairs
originating from decaying sneutrinos and neutralinos.
%
%
%

In the following
the partial decay widths of sneutrinos are calculated for the
first time under
the assumption that there are no two-body decay channels 
kinematically accessible ({\it i.e.} the gauginos are heavier
than all six sneutrino states). 
In the opposite case the final decay products 
are the same but the total decay width should be much bigger.

The parameters are chosen in
accordance with the results of \cite{cdm} in order to assure
the viability of sneutrino CDM. Furthermore, it is assumed that
the $e$- and $\mu$-sneutrinos have masses between the values
of the heavy and light $\tau$-sneutrinos, if they are heavier
additional decay channels into the heavier $\tau$-sneutrino
are present.    

\begin{table} [t]
\begin{center}
\begin{math}
\begin{array}{|l|ccl|l|}
\hline
\mbox{Leptonic:}\;\; & \tilde{\nu}_{\tau}^h 
                     & \ra 
                     & \tau^{\pm} \tau^{\mp} E \hspace{-2mm}/\;\; 
                     & \chi^{\pm}, Z^0  \\
                     & \tilde{\nu}_{\tau}^h
                     & \ra 
                     & l^{\pm} l^{\mp} E \hspace{-2mm}/\;\; (l=e,\mu) 
                     & Z^0  \\
                     & \tilde{\nu}_{\tau}^h 
                     & \ra 
                     & \tau^{\pm} \tau^{\pm} l^{\pm} l^{\pm} 
                       E \hspace{-2mm}/\;\; (l=e,\mu)\;\; 
                     & \chi^{\pm} \\
                     & \tilde{\nu}_l 
                     & \ra 
                     &\tau^{\pm} l^{\mp}
                     E \hspace{-2mm}/\;\; (l=e,\mu) \;\; 
                     & \chi^{\pm}  \\ \hline
\mbox{Hadronic:}\;\; & \tilde{\nu}_{\tau}^h 
                     & \ra 
                     & q \overline{q} E \hspace{-2mm}/\;\;\;\ 
                     & Z^0  \\ \hline
\end{array}
\end{math}
\caption{Visible sneutrino decay channels into the LSP-sneutrino 
$\tilde{\nu}_{\tau}^l$ by exchange of $\chi^{\pm}$ and $Z^0$.}
\label{signatures}
\end{center}
\end{table} 


All sneutrinos but the LSP
decay via neutralino, chargino and
$Z^0$ three-body decays into the lighter $\tau$-sneutrino plus fermions
(note that there is no Higgs-mediated channel since the 
Higgs couples only either to two light or two heavy sneutrinos).
These decay modes are depicted in figure \ref{feynhl}. 
The decay width of the decay $\tilde{\nu}_m \ra \tilde{\nu}_n +
two\;fermions$ is
\be \label{width}
\Gamma_{\mbox{tot}}=
\frac{1}{2 m_m} \frac{1}{(2 \pi)^5} \frac{\pi^2}{4 m_m^2} g^4
(\Gamma'_{\tilde{\chi}^0} + 
 \Gamma'_{\tilde{\chi}^+} + 
 \Gamma'_{Z^0} + 
 \Gamma'_{\tilde{\chi^0 Z^0}} + \Gamma'_{\tilde{\chi}^+ Z^0})\; . 
\ee
Note that there is no interference between chargino and neutralino
contributions for any final state. The widths of the intermediate
states have been neglected, propagator factors $1/(M^2-X)$
are denoted as $P(M,X)$ and $c_W$ is shorthand for $\cos \Theta_W$
etc. The conventions for the gaugino mixing matrices follow 
\cite{susy}. The contribution of the $Z^0$-channel is
\be
\Gamma'_{Z^0} & = & \frac{2}{c_W^2} \int d s_2 |P(m_{Z^0},s_2)|^2
              B_{Z^0}(\;(|O_L^f|^2+|O_R^f|^2)\,F+
               \mbox{Re}[O_L^f O_R^{f *}]\,G\;)  \\
F &=&  - \frac{1}{3} (3 A_{Z^0}^2 + B_{Z^0}^2) 
       + A_{Z^0} (m_m^2 + m_n^2 + 2 m_f^2 - s_2) 
       - m_f^2\frac{(\Delta \tilde{m})^4}{m_{Z^0}^2} \nn \\
  &+&  (m_m^2+m_f^2-\frac{1}{2}s_2)(\frac{1}{2}s_2-m_f^2-m_n^2)
       - \frac{1}{4}(s_2-m_f^2)(2m_m^2+2m_n^2-s_2) \nn \\
G & =& m_f^2
       (2 \frac{(\Delta \tilde{m})^4}{m_{Z^0}^2}-2m_m^2-2m_n^2+s_2)
\nn
\ee
Here we have defined 
\be
A_{Z^0} =   m_f^2 + \frac{1}{2}(m_m^2 + m_n^2 - s_2);\;\;
B_{Z^0} =  \frac{1}{2 s_2} \lambda^{1/2}(s_2,m_m^2,m_n^2)
              \sqrt{s_2^2 - 4 m_m^2 m_f^2} \nn
\ee \vspace{-0.6cm}
\be
(\Delta \tilde{m})^2=m_m^2-m_n^2;\;\;
O_f^L=T_{3f}-e_f s_W^2;\;\;O_f^R=-e_f s^2_W \nn
\ee
and the integration range is $(s_2)_{min,max}$=$4 m_f^2,(m_m - m_n)^2$.

The contribution of the chargino-mediated channel is
\be 
\Gamma'_{\tilde{\chi}^+} & = &
    \sum_{j,j'} |V_{j1}|^2 |V_{j'1}|^2 \nn \\
&\times & \int d s_2 P(m_j,s_2) 
    [ P(m_{j'},s_2) 4 B_{f \ovl{f}'}
      (s_2 - m_n^2 + m_f^2)(m_m^2 - s_2 - m_{\ovl{f}'}^2) \nn \\
&-&  2 B_{f \ovl{f}'} s_2 + 
      (m_f^2 m_{\ovl{f}'}^2 - m_m^2 m_n^2 + m_{j'}^2 s_2)
      \ln \frac{A_{f \ovl{f}'} + B_{f \ovl{f}'} + m_{j'}^2 - X}
      {A_{f \ovl{f}'} - B_{f \ovl{f}'} + m_{j'}^2 - X} ] \nn \\ 
&+&  \;\; (m_f \longleftrightarrow m_{\ovl{f}'})
\ee
Here we have defined
\be \label{affbff}
A_{f \ovl{f}'} & = & m_f^2 + m_{\ovl{f}'} 
    - \frac{1}{2 s_2} (s_2+m_f^2-m_n^2)(s_2-m_m^2+m_{\ovl{f}'}^2) \nn \\
B_{f \ovl{f}'} & = & \frac{1}{2 s_2} 
                    \lambda^{1/2}(s_2,m_f^2,m_n^2)
                    \lambda^{1/2}(s_2,m_{\ovl{f}'}^2,m_m^2)\\
X & = & m_m^2 + m_f^2 + m_{\ovl{f}'}^2 + m_n^2 - s_2 \nn
\ee  
The integration range is 
$(s_2)_{min,max}$=$(m_f + m_n)^2,(m_m - m_{\ovl{f}'})^2)$.
Here and in the following $\lambda$ is defined as usual as
$\lambda(a,b,c)=a^2+b^2+c^2-2ab-2ac-2bc$ and $V$ is the chargino
mixing matrix.


The contribution of neutralino-mediated decays into the final
states $\nu \nu$, $\ovl{\nu} \nu$, $\ovl{\nu} \ovl{\nu}$ is
\be
\Gamma'_{\tilde{\chi}^0} & = &  
         \Gamma^{' \nu \nu}_{\tilde{\chi}^0} + 
         \Gamma^{' \ovl{\nu}\;\ovl{\nu}}_{\tilde{\chi}^0} + 
         \Gamma^{' \ovl{\nu} \nu}_{\tilde{\chi}^0} \\
\Gamma^{' \nu \nu}_{\tilde{\chi}^0} & = & 
       \frac{4}{2 !} \sum_{i,j} (O_i^2 O_j^{* 2} + O_i^{* 2} O_j^2)
       \int 
       \frac{d s_2}{(m_i^2-s_2) (m_j^2-s_2)}
       \frac{1}{2 s_2} (m_m^2 - s_2)^2 (s_2 - m_n^2)^2 \nn \\
& = & \Gamma^{' \ovl{\nu}\;\ovl{\nu}}_{\tilde{\chi}^0} \nn \\ 
\Gamma^{' \ovl{\nu} \nu}_{\tilde{\chi}^0} & = &
       8 \sum_{i,j}|O_i|^2 |O_j|^2
       \int d s_2 P(m_i,s_2)
       [\frac{1}{2 s_2} P(m_j,s_2) (m_m^2-s_2)^2 (s_2-m_n^2)^2 \nn \\
&+&  (m_n^2-s_2)(m_m^2-s_2) +
       (m_m^2 m_n^2-m_j^2 s_2) \ln 
       \frac{s_2+m_j^2-m_m^2-m_n^2}{m_j^2-m_m^2 m_n^2/s_2}] \nn
\ee
The integration limits are $(s_2)_{min,max}$=$m_n^2,m_m^2$
and we have defined ($N$ is the neutralino mixing matrix)
$O_i = \frac{1}{2}(N_{i1} - t_W N_{i2})$.

The interference between chargino- and $Z^0$-contributions is
\be
\Gamma'_{\tilde{\chi}^+ Z^0}  
  & = &
    \frac{8}{c^2_W} \sum_j |V_{j1}|^2 
    \int 
    d s_2 P(m_j,s_2)[O^f_L 2 s_2 B_{ff} F + O_R^f \frac{m_f^2}{2} G] \\
F & = & [s_2  m_Z^2 - s_2 (m_m^2 - s_2) -
      \frac{m_n^2}{2} (m_f^2 - 2 m_m^2 + 2 s_2) -
      \frac{m_f^2}{2} (m_m^2 + 2 s_2) \nn \\
  & + &  \frac{1}{2} \frac{(\Delta \tilde{m})^2}{m_Z^2} m_f^2 
      (m_m^2 - m_n^2)] 
      \ln \frac{ A_{ff} + B_{ff} - m_Z^2}{A_{ff} - B_{ff} - m_Z^2}\nn \\
G & = & \frac{1}{2} m_f^2 [2 s_2 - 2 m_f^2 + m_m^2 + m_n^2 -
      \frac{(\Delta \tilde{m})^2}{m_Z^2}(m_m^2 - m_n^2)]
      \ln \frac{A_{ff} + B_{ff} - m_Z^2}{A_{ff} - B_{ff} - m_Z^2}. \nn
\ee 
This term is only relevant for the transition 
$\tilde{\nu}^{\tau}_1 \ra \tilde{\nu}^{\tau}_2 \tau \ovl{\tau}$.
The integration limits are 
$(s_2)_{min,max}$=$(m_f + m_n)^2,(m_m - m_f)^2$ and the functions 
$A_{ff},B_{ff}$ have been defined in (\ref{affbff}). 
\newpage

\begin{center}
\begin{tabular}{|l|c||c|c|c||c|c|c||l|}
\hline
          & $\Gamma_{\chi^-}$ & & $\Gamma_{\chi^0}$ & $m_{\chi^0}$ & 
                         & $\Gamma_{\chi^0}$ & $m_{\chi^0}$ & \\ \hline
$M_2=400$ &  -  &        
          &  -  & 77.1 & 
          &  -  &69.9& Ia \\ \cline{2-2} \cline{4-5}\cline{7-9}
          &0.133&    &1.217&86.9& &23.20&82.5&Ib   
                            \\ \cline{2-2}\cline{4-5}\cline{7-9}
          &0.032&    &0.248&97.6& &2.165&96.6&IIa
                            \\ \cline{2-2}\cline{4-5}\cline{7-9}
          &0.114&    &0.943&88.2& &14.29&84.1&IIb
                            \\ \cline{1-2}\cline{4-5}\cline{7-9}
                               \cline{1-2}\cline{4-5}\cline{7-9}
$M_2=800$ &0.009& \raisebox{0.25cm}[-0.25cm]{A} 
          &0.094&89.5& 
                  \raisebox{0.25cm}[-0.25cm]{B}
          &1.036&86.5&Ia  \\ \cline{2-2}\cline{4-5}\cline{7-9}
          &0.006&    &0.041&94.0& 
          &0.348&92.3&Ib  \\ \cline{2-2}\cline{4-5}\cline{7-9}
          &0.003&    &0.021&98.9& 
          &0.152&98.5&IIa \\ \cline{2-2}\cline{4-5}\cline{7-9}
          &0.006&    &0.038&94.6& 
          &0.311&93.0&IIb \\ \hline \hline
$M_2=-400$&0.032&      
          &3.480&95.1& 
                                        & - &74.5&Ia
                            \\ \cline{2-2}\cline{4-5}\cline{7-9}
          &0.114&     &2.354&93.5& &185.2&82.1&Ib
                            \\ \cline{2-2}\cline{4-5}\cline{7-9}
          &0.540&     &1.418&87.7& &19.30&89.8&IIb
                            \\ \cline{2-2}\cline{4-5}\cline{7-9}
          &0.133&     &2.227&92.8& &126.1&83.4&IIb
                            \\ \cline{1-2}\cline{4-5}\cline{7-9}
                               \cline{1-2}\cline{4-5}\cline{7-9}
$M_2=-600$&0.008& \raisebox{0.25cm}[-0.25cm]{C}    
         &0.411&99.2& 
                  \raisebox{0.25cm}[-0.25cm]{D}
         &10.09&89.3&Ia \\ \cline{2-2}\cline{4-5}\cline{7-9}
         &0.020&     &0.331&96.4& &49.52&93.4&Ib
                            \\ \cline{2-2}\cline{4-5}\cline{7-9}
         &0.039&     &0.250&93.1& &2.334&94.9&IIa
                            \\ \cline{2-2}\cline{4-5}\cline{7-9}
          &0.022&     &0.321&96.0& &4.522&94.0&IIb
                            \\ \hline
\end{tabular}
\end{center}

\noindent
Table 2: Partial decay widths in $eV$ for the transitions 
$\tilde{\nu}_{e,\mu} \ra \tilde{\nu}_{\tau}^{LSP}$ when 
two-body transitions are kinematically excluded. 
The mass splitting of the 
$\tau$-sneutrinos is taken to be $10GeV$, the mass of 
$\tilde{\nu}_{\tau}^{LSP}$ is $70GeV$, the masses of the $e$- and
$\mu$-sneutrinos are $75GeV$ and their mass-splitting has been
neglected. 
The different ratios
are $M_1=M_2$ (A), $M_1=(5\alpha_Y/3\alpha_W) M_2$ (B), 
$M_1=-(5\alpha_Y/3\alpha_W) M_2$ (C) and 
$M_1=-(\alpha_Y/\alpha_W) M_2$ (D).
``I'' (``II'') stands for 
$\mu$=100 (-100) $GeV$, ``a'' (``b'') stands for $\tan\beta$=2 (40). 
$M_2$ is given in $GeV$.
The parameters and in particular the ratios $M_1/M_2$ 
have been chosen in accordance with \cite{cdm}:
for these parameters $\tilde{\nu}_{\tau}^{LSP}$ has a relic 
abundance sufficient to account for the CDM.
$\Gamma_{\chi^-}$ ($\Gamma_{\chi^0}$) denotes the 
chargino- (neutrali\-no-) mediated channel. 
If the $e$- and $\mu$-sneutrinos are heavier than
$\tilde{\nu}_{\tau}^h$ an additional decay channel of the former
into the latter is open. 
The cases where the lightest neutralino is 
lighter than the decaying sneutrinos have been omitted.

\vspace{0.8cm}

Finally, the interference between neutralino and $Z^0$ contributions:
\be
\Gamma'_{\tilde{\chi}^0 Z^0}  
  & = &
       \frac{8}{c^2_W} \sum_i |O_i|^2 O_{\nu}^L
       \int 
       d s_2 P(m_i,s_2) 
       [(m_m^2-s_2)(s_2-m_n^2) \\
  & + & (s_2 m_{Z^0}^2-m_n^2(s_2-m_m^2) + s_2(s_2-m_m^2)) 
       \ln (1-\frac{(s_2-m_m^2)(s_2-m_n^2)}{m_{Z^0}^2 s_2})] \nn
\ee
The integration limits are $(s_2)_{min,max}$=$m_n^2,m_m^2$ and
this term is relevant only for the $\nu \ovl\nu$ final state.


\newpage

\begin{center}
\begin{tabular}{|l|c|c||c|c||c|c||c|}
\hline
          &   $\Gamma_{e\tau}$ & $\Gamma_{\tau\tau}$ 
          & & $\Gamma_{tot}$   
          & & $\Gamma_{tot}$ & 
          \\ \hline
$M_2=400$ &  -  &  -  &     &  -  &     &  -  & Ia
          \\ \cline{2-3}\cline{5-5}\cline{7-8}
          &0.195&0.956&  &47.261&   &84.052&Ib
          \\ \cline{2-3}\cline{5-5}\cline{7-8}
          &0.037&0.166&  &42.748&   &48.603&IIa
          \\ \cline{2-3}\cline{5-5}\cline{7-8} 
          &0.160&0.613&  &46.043&    &70.691&IIb
          \\ \cline{1-3}\cline{5-5}\cline{7-8}
$M_2=800$ &0.013&0.777
          &\raisebox{0.25cm}[-0.25cm]{A}
          &42.379
          &\raisebox{0.25cm}[-0.25cm]{B}
          &46.191&Ia
          \\ \cline{2-3}\cline{5-5}\cline{7-8}
          &0.009&0.955& &42.083& &43.952&Ib
          \\ \cline{2-3}\cline{5-5}\cline{7-8}
          &0.003&1.359& &42.213&  &43.341&IIa
          \\ \cline{2-3}\cline{5-5}\cline{7-8}
          &0.008&9.931& &42.0797&  &43.819&IIb
          \\ \hline \hline
$M_2=-400$&0.374&0.166 & &51.508&  &   -  &Ia
          \\ \cline{2-3}\cline{5-5}\cline{7-8}
          &0.160&0.613 & &49.738&  &280.64&Ib
          \\ \cline{2-3}\cline{5-5}\cline{7-8}
          &1.917&18.315& &68.619&  &99.441&IIa
          \\ \cline{2-3}\cline{5-5}\cline{7-8}
          &0.195&0.956 & &49.853&  &212.69&IIb
          \\ \cline{1-3}\cline{5-5}\cline{7-8}          
$M_2=-600$&0.009&0.895          
          &\raisebox{0.25cm}[-0.25cm]{C}
          &44.163
          &\raisebox{0.25cm}[-0.25cm]{D}
          &64.086&Ia
          \\ \cline{2-3}\cline{5-5}\cline{7-8}
          &0.027&0.362& &43.317& &56.409&Ib
          \\ \cline{2-3}\cline{5-5}\cline{7-8}
          &0.068&0.078&  &42.728& &48.970&IIa
          \\ \cline{2-3}\cline{5-5}\cline{7-8}
          &0.030&0.309&  &43.225& &53.735&IIb
          \\ \hline
\end{tabular}
\end{center}

\noindent
Table 3: Partial decay widths $\Gamma_{e\tau,\tau\tau}$ into final 
states containing $\tau$'s in $eV$ for the transitions 
$\tilde{\nu}_{\tau}^h \ra \tilde{\nu}_{\tau}^{LSP}$ 
for the parameters defined in table 2. 
$\Gamma_{e\tau}$ denotes the (chargino-mediated) width into a
$e^{\pm}\tau^{\mp}$-pair ($\Gamma_{\mu\tau}\approx\Gamma_{e\tau}$), 
$\Gamma_{\tau\tau}$ is the
(chargino- and $Z^0$-mediated) width containing 
$\tau^{\pm}\tau^{\mp}$-pairs in the final state and $\Gamma_{tot}$
is the total width (when two-body transitions are kinematically 
excluded). The contribution
of the $Z^0$-channel is constant and in particular the width
into the remaining visible hadronic and fermionic states is
$\Gamma_{other}=21.850eV$. The total visible width is 
$\Gamma_{other}+\Gamma_{\tau\tau}+\Gamma_{e\tau}+\Gamma_{\mu\tau}
\approx 22eV-23eV$.

\vspace{8mm}
\noindent

Provided the escaping fermions being not too soft to be
detected the expected signatures are listed in table 
\ref{signatures}. The visible channels consist of one or 
more lepton pairs or jets and missing momentum. Each 
decaying sneutrino possesses individual characteristics.
The transitions $\tilde{\nu}_{e,\mu}$ are background-free
on tree-level, whereas the transition 
$\tilde{\nu}_{\tau}^h \ra \tilde{\nu}^{LSP}_{\tau} 
\tau^{\pm}\tau^{\mp} + E \hspace{-2mm}/ \;$ is simulated
by the decay of a heavier into a lighter neutralino which
subsequently decays into $\tilde{\nu}_{\tau}^{LSP} \nu$.
However, in this case the leptons possess another angular 
distribution. Hence, for $\tilde{\nu}_{\tau}^2$ being the LSP 
initially produced sneutrinos result in a signal which allows 
for each visible event to track the flavour of the decaying 
sneutrino, deduce its mass using missing momentum and therefore 
pin down their properties {\it e.g.} in 
$e^+ e^- \ra \tilde{\nu}_l^1 \tilde{\nu}_l^2$. 
%
%
For the example parameters used in tables 2,3
in the case of decaying heavy $\tau$-sneutrinos the branching 
ratio of the visible decay channels is about 50\%, the rest being
invisible neutrino final states from neutralino- and $Z^0$-exchange.
The contribution of the $Z^0$-channel is constant and in particular 
the width into the remaining visible hadronic and fermionic states is
$\Gamma_{other}=21.850eV$. The total visible width is 
($\Gamma_{e\tau}\approx\Gamma_{\mu\tau}$)
\be 
\Gamma_{other}+\Gamma_{\tau\tau}+\cdot\Gamma_{e\tau}+\Gamma_{\mu\tau}
\approx 22eV-23eV \; .\nn
\ee
Due to destructive 
interference of the chargino- and $Z^0$-mediated channels for the
parameters chosen the $\tau\tau$-final state typically makes up 
only a few percent of the visible channels. An exception is
the case C/D, $M_2=-400$, IIa. The $l\tau$-channel accounts
roughly for 0.1\% to 1\% of the visible events. About 10\% of the
$\tilde{\nu}^l$ decay themselves into the visible state 
(see table 2). 
Hence, for the projected 
NLC-luminosity of 50 $(fb \cdot y)^{-1}$ \cite{nlc} and a 
sneutrino pair production cross-section of ${\cal O}(100 fb)$ 
(for $m_{\tilde{\nu}} \sim {\cal O}(100GeV))$ at most a few
events containing a single-sided two-acoplanar $l^{\pm}\tau^{\mp}$-pair 
is expected per year. 

%
%

The situation is different for the decays of $e$- and 
$\mu$-sneutrinos. The $Z^0$-channel does not contribute 
(in the absence of mixing in generation space) and 
$\Gamma_{vis}$=$\Gamma_{l \tau}$. For most of the parameters
relevant for sneutrino CDM the visible partial decay width is 
of order 10\% so that for the integrated luminositity 50 
$(fb \cdot y)^{-1}$ and a sneutrino pair production cross-section 
of ${\cal O}(100 fb)$  (for $m_{\tilde{\nu}} \sim {\cal O}(100GeV))$  
several hundred visible sneutrino decay events containing
a $\tau^{\pm} l^{\mp}$-pair are expected. Such a signal
would be a clear hint for $\tilde{\nu}^{LSP}_{\tau}$.
On the other hand, in the case $M_1=M_2/2$ for high $M_2$ 
the visible width is negligible: $\Gamma_{vis}\approx 10^{-10}eV$,
$\Gamma_{\chi^0}\approx 10^{-2}eV$.

\vspace*{0.5cm}

In conclusion, the LSP being a sneutrino changes the signals 
expected for the decays of sparticles considerably in comparison
with the case of a neutralino LSP. For the cosmologically 
interesting case of sneutrino CDM sneutrinos have to violate 
$L$ and only the lighter $\tau$-sneutrino can account for the
CDM. In this case final sparticle-decay states are expected to 
frequently contain generation-diagonal fermion-pairs (which are 
present for a neutralino LSP as well) and in particular 
$\tau^{\pm} l^{\mp}$-pairs ($l$=$e$, $\mu$) which would provide 
a clear signal for sneutrino-LSP. The latter ones are not present 
for a neutralino LSP. In the case that all sneutrino states are 
lighter than the neutralinos in most of the range of interest for 
sneutrino CDM the partial decay widths into the visible final states 
are of order 10\% of the total decay width for decaying 
$\tilde{\nu}^{e,\mu}$ and 50\% for decaying $\tilde{\nu}^h_{\tau}$ of 
the total decay width producing hundreds of events at {\it e.g.} the
NLC. It therefore should be possible to settle the question of the 
sneutrino being CDM at such a facility.

\vspace{0.5cm}
\centerline
{\bf Acknowledgements}

\vspace{0.3cm}
\noindent
St. K. wants to thank INFN Sezione Perugia for hospitality. 
M.H. is 
supported by the Spanish DGICYT under grant PB95-1077 and by the
European Union's TMR program under grants ERBFMRXCT960090 and
ERBFMBICT983000.


\end{document}